# Spin-1/2 antiferromagnets in 2 dimensions

C. Lhuillier and P. Sindzingre

*Laboratoire de Physique Théorique des Liquides, Université Pierre et Marie Curie, case 121, 4 Place Jussieu, 75252 Paris Cedex, France.*

The properties of the ground-state and first excitations of 2-dimensionnal quantum antiferromagnets are rapidly sketched. A special emphasis is put on the gapped phases: Valence Bond Crystals and the two types of Spin Liquids. New results on the spin susceptibility of small samples of the Heisenberg antiferromagnet on a kagomé lattice are displayed for comparison with Kagemaya *et al.* experimental results (this conference).

## 1.INTRODUCTION

In two dimensions the T=0 phase of an SU(2) invariant hamiltonian can break SU(2) invariance or not. The SU(2) symmetry breaking phases are Néel-like phases with 2 or 3 sublattices and a reduced magnetization. These phases are the ground-states of the Heisenberg model on the square, hexagonal and triangular lattices and in a large range of parameters for the J1-J2 models.

T=0 phases which do not break rotational spin symmetry have a spin gap (except at critical points). But in two dimensions, contrary to the one-dimensionnal case, these spin gapped phases can exhibit various physical behaviors. Through the analysis of exact spectra of small samples we have characterized numerically three kinds of generic phases with a spin gap: the Valence Bond Crystals and the Resonating Valence Bond Spin Liquids (type I and II).

## 2.THE VALENCE BOND CRYSTALS

They are the two-dimensional analogs of the dimerized phase appearing in the J1-J2 chain (for J2/J1 > 0.24) or in models with Spin Peierls instability. They have long range order in dimer-dimer correlations (4-point spin correlations), or in higher order spin-spin correlations. Various kinds of Valence Bond Crystals are sketched in Fig.1. Fig.1-a is a columnar Valence Bond Crystal: it is the ground-state of the J1-J2 model on the hexagonal lattice (for J2/J1 ≅0.4) [1], and possibly on the square lattice (for J2/J1 ≅0.5) [2,3,4]. Up to now staggered configurations of type 1-b have only been seen in toy models [5]. Configuration 1-c is the well known ground-state of the Shastry-Sutherland model [6]: a very good experimental realization of this model is obtained in $SrCu_2(BO_3)_2$[7]. Configuration 1-d is a more exotic VBC ground-state, where there is long range order in 4-spin S=0 plaquettes (bold lines of Fig. 1-d). It is the ground-state observed in the Heisenberg model on the two-dimensional pyrochlore: a two dimensional lattice with corner sharing tetrahedrons where all couplings are identical. The classical model on such a lattice is disordered and has an infinite local degeneracy (all configurations which have a spin 0 on each tetrahedron)[8]. The quantum spin-1/2 magnet is quite different: 4 spins around a void cooperate to form a singlet state and the system has long range order in this 4-spin observable[9].

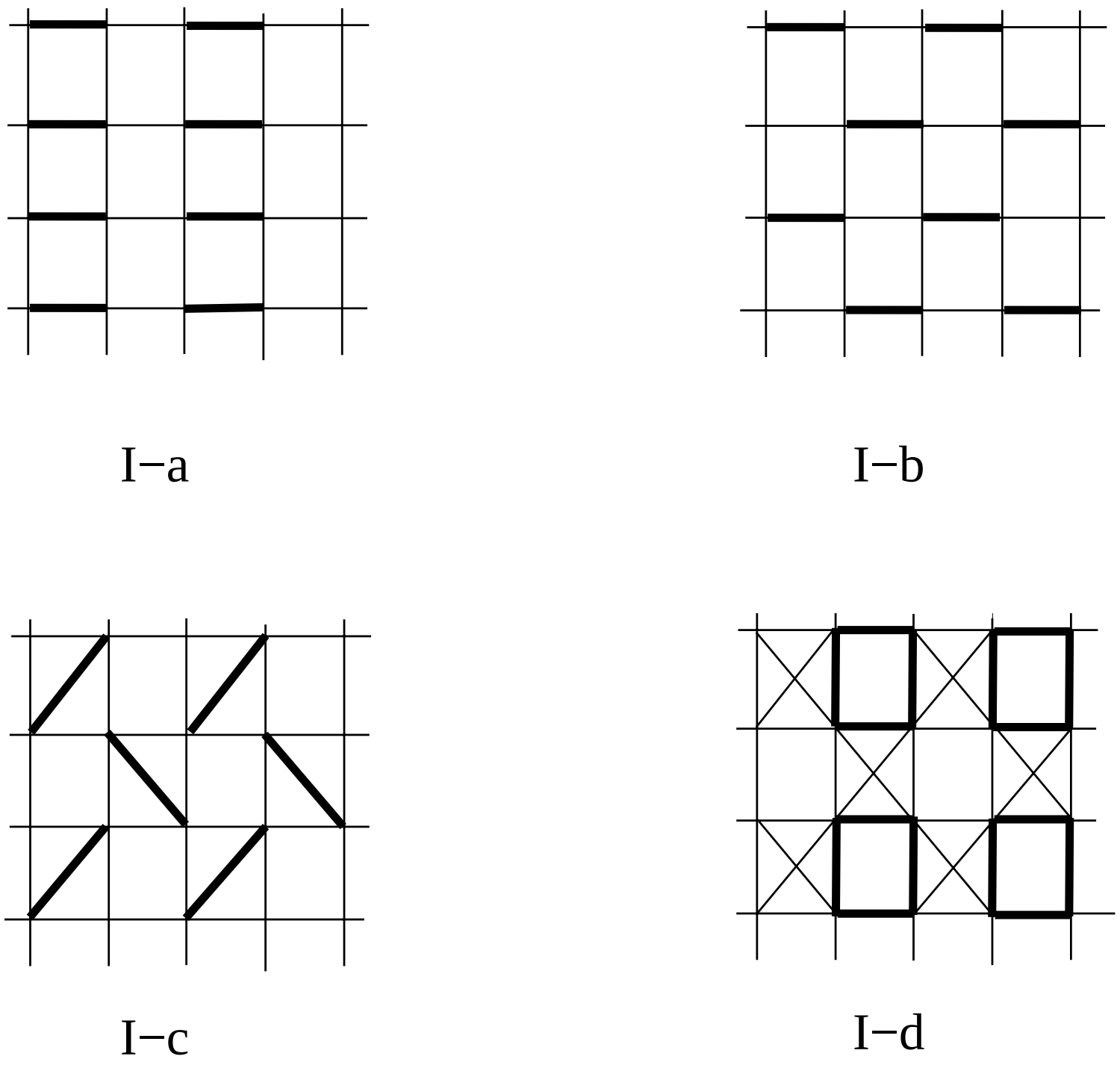

Fig. 1 Some configurations of Valence Bond Crystals.

The first excitations of all these models can in general be described as propagating modes (more or less dispersive) with integer spins.

## 3.THE RESONATING VALENCE BOND SPIN LIQUIDS

In first approximation the Valence Bond Crystal ground-state can be described with *specific* Short Range Valence Bond configurations displaying long range order as those depicted in Fig.1. In some situations and specifically on the triangular lattice, disordered SRRVB configurations could have energy quite comparable to the ordered ones. In those situations *quantum resonances between all the short range configurations* can lead to a ground-state without any long range order in dimer nor in larger S=0-plaquette states: it is a Resonating Valence Bond Spin Liquid (RVBSL)[10].

The second important difference between VBC and RVBSL concerns excitations. Excitations can be created by the promotion of a dimer (or a more complex S=0 plaquette) to a triplet state (or excited S=0 or S=2 states in more complex cases). In the VBC case two spins ½ of a triplet cannot separate far apart. Such a process would create a string of misaligned valence bonds (see Fig.2), with an energy increasing as the length of the string. This process is energetically unfavourable and the two spins ½ cannot separate from each other (the spinons are confined). On the other hand, as any disordered configuration is equally probable in the RVBSL, the separation of two spins ½ suffers no energetic constraint and spinons in the RVBSL are deconfined. The first expected consequence of this property is that excitations in the RVBSL should appear as continua and not as well separated modes.

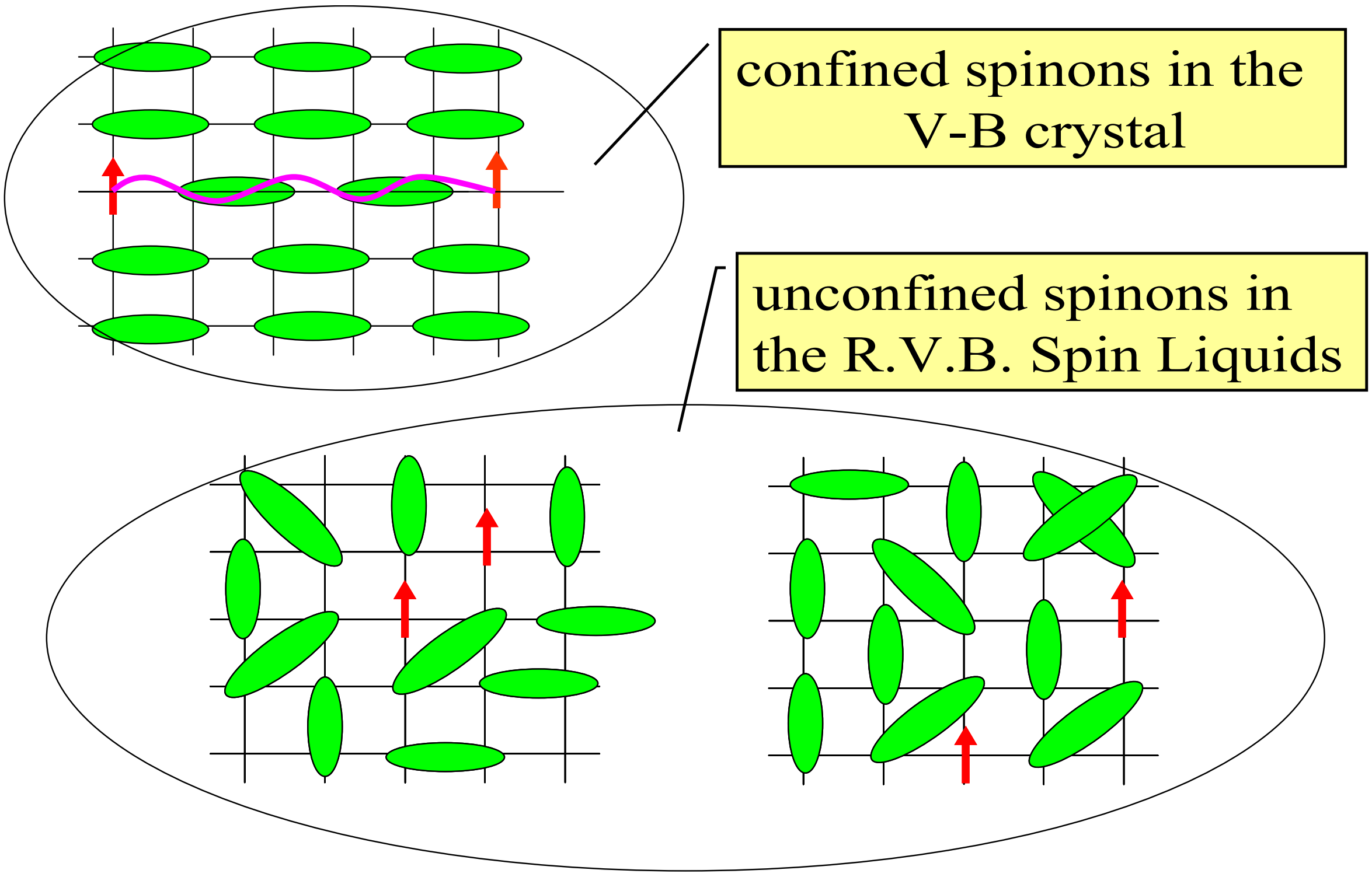

Fig. 2 A sketch of the main difference between excitations in VBC and RVBSL.

All RVBSL are characterized by a downward shift of the spectral density of states: i. e. at a temperature T< 0.5 $\theta_{CW}$ (where $\theta_{CW}$ is defined from the high temperature behavior of the specific heat) only half of the degrees of freedom are frozen. By the fact these systems display two scales of temperature as was remarked by Ramirez, which takes this fact as the measurement of the degree of frustration of these antiferromagnets[11]. A theoretical illustration of this property is given in Fig. 3. The first curve is the specific heat of the pure Heisenberg model[⊥] on the triangular lattice: this system is in a Néel state and the specific heat shows only one temperature scale of the order of the coupling constant. The two other curves are specific of Resonating Valence Bond Spin Liquids (type I and II): both of these curves show a second maximum at a temperature much lower than the main coupling constant and approximately half of the entropy is developed in this low temperature range. In this family of RVBSL we have observed two different classes of antiferromagnets: both have a gap to magnetic excitations but type I RVBSL have a gap in the singlet sector whereas type II RVBSL display a continuum of singlets adjacent to the ground-state.

**3.1 Type I Resonating Valence Bond Spin Liquid**

Up to now we have observed this behavior in two models:

i) the Multi-Spin Exchange model on the triangular lattice[12]
ii) the J1-J2 model on the hexagonal lattice[1] (with ferromagnetic J1 coupling, antiferromagnetic J2 and J2/J1 ≅ - 0.25).

---

[⊥]In this paper the Heisenberg Hamiltonian is defined as: $H = J \sum S_i.S_j$ , where the sum runs on pairs of next neighbours.

In both cases there is a strong competition between interactions, the valence bond coverings prefer a triangular lattice and the next neighbor correlations are extremely weak and sometimes weakly ferromagnetic.

- Néel ordered A.F. on the triang. latt.

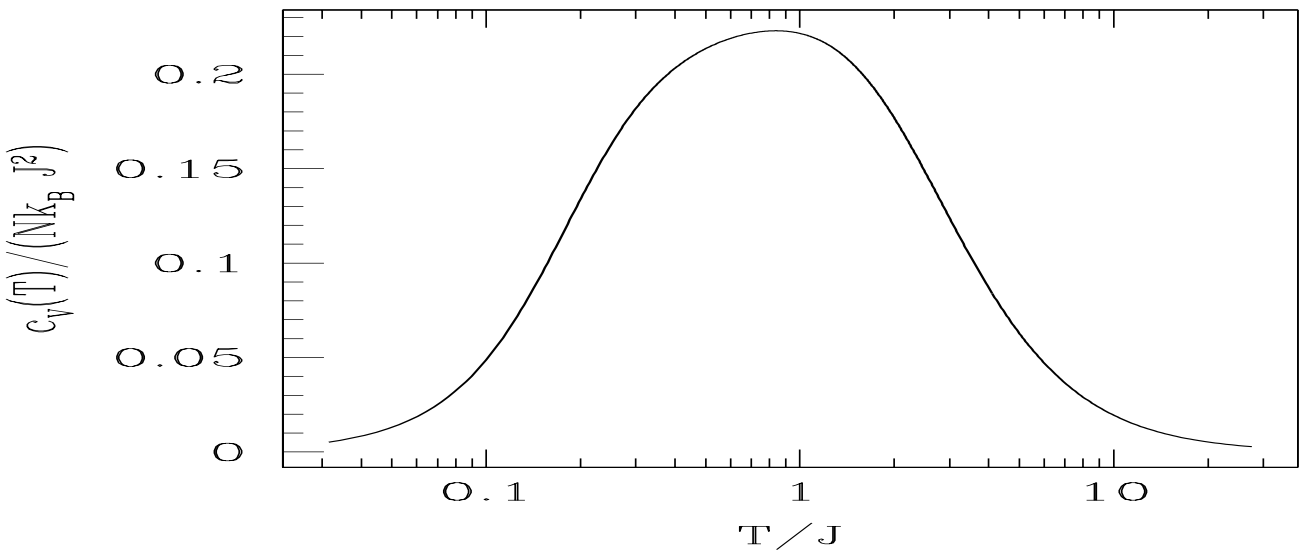

- M.S.E. Spin Liquid on the triang. latt.

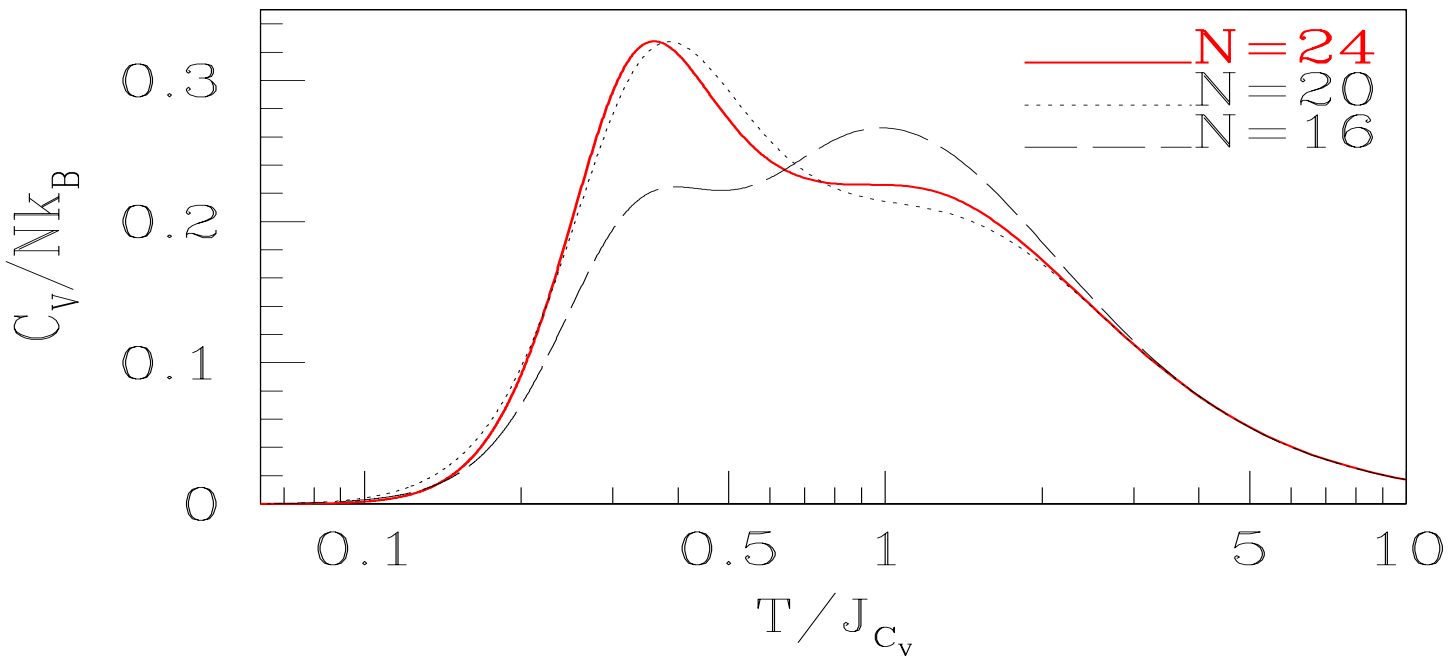

- Spin Liquid on the kagomé lattice

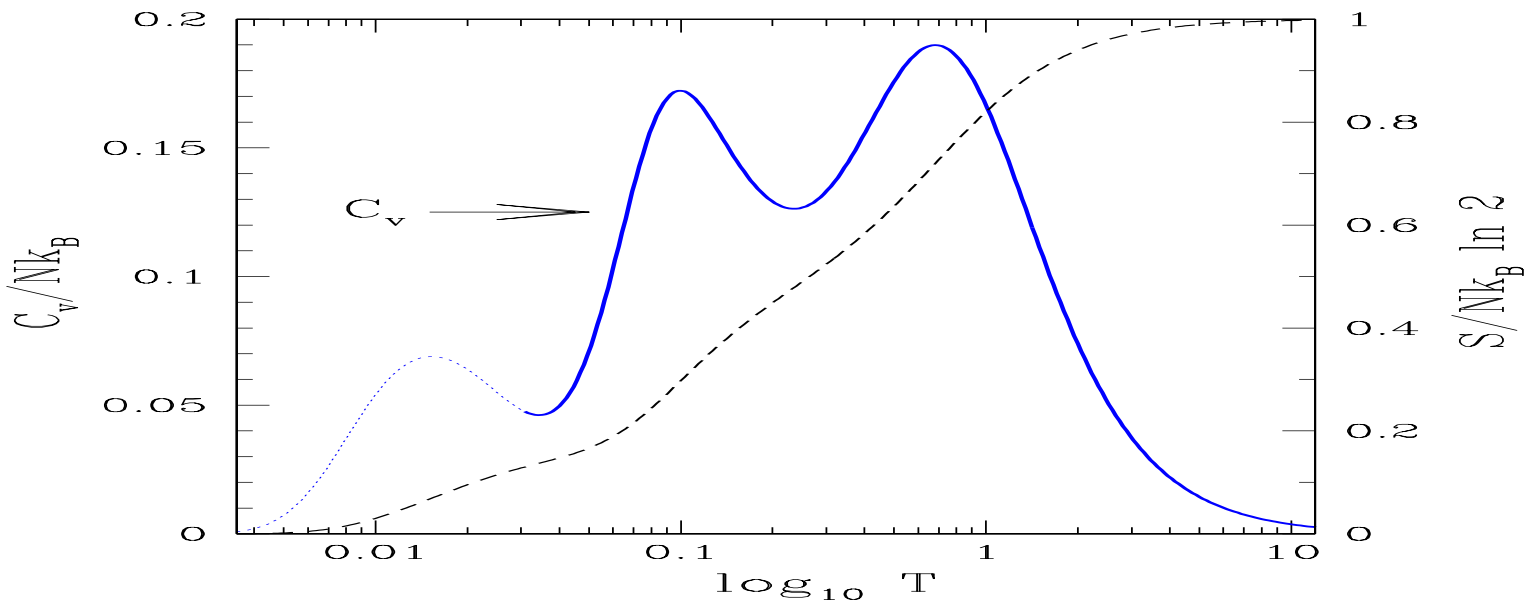

Fig. 3. Illustration on the specific heat of the downward shift of the density of states. The upper curve is the specific heat of the Heisenberg model on the triangular lattice: a T=0 Néel ordered phase; the middle curve is the specific heat of the MSE model of ref [12]: a type I Spin Liquid; the lower curve describes the specific heat and the entropy of the Heisenberg model on the kagomé lattice: a type II Spin Liquid (in this last case the third peak at a very low temperature is an unphysical finite size effect).

Specific to the type I RVBSL are the thermally activated low temperature behavior of both the specific heat and spin susceptibility. The existence of two temperature scales has been clearly seen in the specific heat of the low density phase of $^3$He on graphite (supposed to be a type I RVBSL)[13] and some indications of the existence of a spin gap of the order of 50 μK have recently been observed on the spin susceptibility[14].

These RVBSL have interesting topological properties [15]. A thorough study of the excitations of this type I RVBSL remains to be done.

### 3.2 Type II Resonating Valence Bond Spin Liquid

An unexpected second type of RVBSL has been found some years ago in the pure Heisenberg model on the kagomé lattice[16,17]. On the basis of the classical analysis, which shows an infinite local degeneracy, it was expected that the quantum ground state of this model could be "disordered" *. In fact the ground state of the Heisenberg model on the kagomé lattice is indeed a RVBSL. But it appears that the degeneracy between the various disordered configurations is only marginally lifted by quantum resonances $^+$.

In some sense the type I RVBSL can be seen as a Valence Bond Crystal where the long range order has "molten" because of frustration, competitive interactions and quantum resonances. The type II RVBSL is different in so much as no preferred local configuration do emerge and there is an exponential number of valence bond configurations which are roughly equally favorable [18,19].

This fact has macroscopic consequences.

1) The density of low lying singlets is unexpectedly high: it has been measured on exact spectra that the number of singlet states in a given finite range of energy (J/20) increases exponentially with the system size [17]. This implies that the system has a T = 0 residual entropy and the specific heat is not thermally activated. As singlet states are insensitive to magnetic field, we have argued that this property was at the origin of the very low sensitivity of the specific heat of SrCrGaO to high magnetic fields [20,21].

2) A high density of low lying S=1 states is equally observed just above the spin gap, and supposed to exist in many other subspaces with higher total spins. These magnetic excitations can be described by a low density of spin-1/2 itinerating randomly in a see of valence bonds [17,18]. This is in qualitative agreement with the picture developed by Uemura and collaborators to explain the anomalous relaxation of muons in SrCrGaO [22] and with the exact diagonalizations of samples with an odd number of sites, which could be seen as systems with one spinon excitation [17].

3) We would like to underline now that this unusually large density of magnetic excitations (just above a tiny gap) could be at the origin of a strong reinforcement (eventually of a divergence?) of the spin susceptibility just above the spin gap [23].

---

*The example of the 2-dimensional pyrochlore shows the limitations of such expectations.

$^+$From the theoretical point of view, it has sometimes been argued that any infinitesimal perturbation could lift this degeneracy. We do not think that this is as simple as it looks. In fact in this system where there is supposedly no local observable with a non-zero value in the thermodynamic limit, it seems that finite perturbations are needed to lift this residual degeneracy [15].

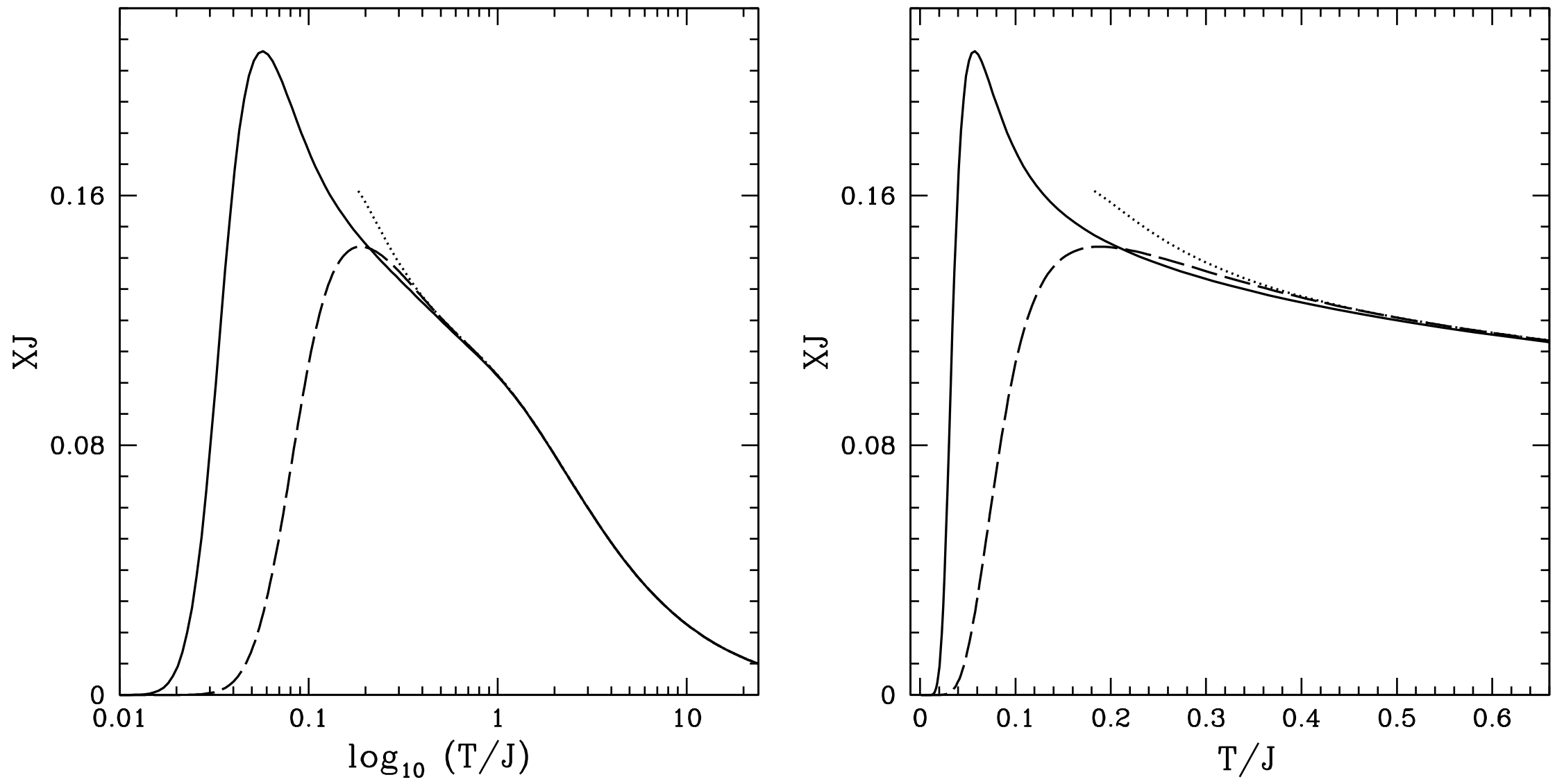

Fig. 4. Spin susceptibility of the Heisenberg model on the kagomé lattice. Dashed line: exact result for an N=18 sample. Solid line: N=36 (this result has been obtained through a maximum entropy procedure from the knowledge of the exact low lying states of the spectrum and of 5 moments of the density of states). Dotted line: Eltsner and Young Padde approximant of the high temperature series [30].

The spin susceptibility of small samples of the Heisenberg nearest neighbor model on the kagomé lattice is displayed in Fig.4. The N=36 sample is still far from the thermodynamic limit (the spin gap $\Delta_{36} = 0.176$ and the extrapolated thermodynamic limit is supposed to be of the order of 0.05). Nevertheless examination of Fig.4 shows the increasing value of the low temperature maximum with sample size. We think that this effect corresponds to a real physical effect directly related to the exponential increase of the density of low-lying magnetic excitations with system size. We cannot exclude a much larger reinforcement of the susceptibility in the thermodynamic limit.

On the other hand such a feature could explain an unusual sensitivity to a very low number of magnetic impurities inducing (anomalous?) spin glass behavior as it has been observed in many compounds SrCrGaO as well as 3-d pyrochlores [24,25,26].

A last question remains to be answered: does this type II RVBSL only appears on “special” lattices (kagomé, square kagomé [27], eventually 3-d pyrochlore[§])? An example of spectrum qualitatively similar to those observed on the kagomé lattice has been found in the multi-spin exchange model on the triangular lattice near-by the 3-sublattice Néel ordered phase [28]. This leads us to suppose that this type II RVBSL could be a generic behavior of strongly frustrated magnets when all couplings favor antiferromagnetism and singlet formation at short range, in contrast to the type I RVBSL where the existence of a competition between near-neighbor ferromagnetic coupling and longer range antiferromagnetic ones seem more favorable.

This short survey does not exhaust the subject of the two-dimensional phases with a gap. A more complete summary could be found in [29] and references therein.

---

[§] In view of the results of exact diagonalizations on N=16 and N=32 samples, it is not excluded that the 3-d isotropic Heisenberg antiferromagnet on a pyrochlore lattice is, in contrast to the 2-d model, a true type II RVBSL.